\documentclass[preprint,12pt]{elsarticle}

\usepackage[colorlinks]{hyperref} 

\usepackage{amsmath}
\usepackage{amssymb}
\usepackage{enumitem}

\usepackage{xargs}
\usepackage[colorinlistoftodos,prependcaption,textsize=tiny]{todonotes}
\newcommandx{\unsure}[2][1=]{\todo[linecolor=red,backgroundcolor=red!25,bordercolor=red,#1]{#2}}
\newcommandx{\change}[2][1=]{\todo[linecolor=blue,backgroundcolor=blue!25,bordercolor=blue,#1]{#2}}
\newcommandx{\info}[2][1=]{\todo[linecolor=olivegreen,backgroundcolor=olivegreen!25,bordercolor=olivegreen,#1]{#2}}
\newcommandx{\improvement}[2][1=]{\todo[linecolor=plum,backgroundcolor=plum!25,bordercolor=plum,#1]{#2}}
\newcommandx{\thiswillnotshow}[2][1=]{\todo[disable,#1]{#2}}
\usepackage{listings}

\definecolor{dkgreen}{rgb}{0,0.6,0}
\definecolor{gray}{rgb}{0.5,0.5,0.5}
\definecolor{mauve}{rgb}{0.58,0,0.8}

\biboptions{}

\newcounter{bla}

\journal{Computer Physics Communications}

\newcommand{\li}{\mathop{\mathrm{Li}}\nolimits}

\hypersetup{colorlinks=true,       
  linkcolor=red,          
  citecolor=green,        
  filecolor=magenta,      
  urlcolor=cyan           
}

\begin{document}

\begin{frontmatter}
    
  \title{
\vskip-1.5cm{\baselineskip14pt\rm
\centerline{\normalsize DESY 16-019\hfill ISSN 0418-9833}
\centerline{\normalsize January 2016\hfill}}
\vskip1.5cm
\texttt{mr}: a \texttt{C++} library for the matching and running of
the Standard Model parameters}
    
  \author[a]{Bernd A. Kniehl\corref{mycorrespondingauthor}}
  \cortext[mycorrespondingauthor]{Corresponding author}
  \ead{kniehl@desy.de}
  \author[a,b]{Andrey F. Pikelner}
  \ead{andrey.pikelner@desy.de}
  \author[a]{Oleg L. Veretin}
  \ead{oleg.veretin@desy.de}
  \address[a]{II.~Institut f\"ur Theoretische Physik, Universit\"at Hamburg,
    Luruper Chaussee 149, 22761 Hamburg, Germany}
  \address[b]{Bogoliubov Laboratory of Theoretical Physics, Joint Institute for
Nuclear Research, 141980~Dubna, Russia}
  
  \begin{abstract}
We present the \texttt{C++} program library \texttt{mr} that allows us to
reliably calculate the values of the running parameters in the Standard Model
at high energy scales. 
The initial conditions are obtained by relating the running parameters in the
$\overline{\mathrm{MS}}$ renormalization scheme to observables at lower
energies with full two-loop precision.
The evolution is then performed in accordance with the renormalization group
equations with full three-loop precision.
Pure QCD corrections to the matching and running are included through four
loops.
We also provide a \texttt{Mathematica} interface for this program library.

  \end{abstract}


\end{frontmatter}

\pagebreak

{\bf PROGRAM SUMMARY}

\begin{small}
  \noindent
  {\em Manuscript title:} \texttt{mr}: a \texttt{C++} library for the matching
and running of the Standard Model parameters\\
  {\em Authors:} B.~A.~Kniehl, A.~F.~Pikelner, O.~L.~Veretin \\
  {\em Program title:} \texttt{mr} \\
  {\em Journal reference:} \\
  {\em Catalogue identifier:} \\
  {\em Licensing provisions:} GPLv3 \\
  {\em Programming language:} \texttt{C++} \\
  {\em Computer:} IBM PC \\
  {\em Operating system:} \texttt{Linux}, \texttt{Mac OS X} \\
  {\em RAM:} 1~GB \\
  {\em Number of processors used:} available with \texttt{OpenMP} \\
  {\em Supplementary material:} 
  {\em Classification:} 11 Elementary Particle Physics,
 11.1 General, High Energy Physics and Computing \\
  {\em External routines/libraries:} \texttt{TSIL}~[1], \texttt{OdeInt}~[2] \\
  {\em Nature of problem:}\\
The running parameters of the Standard Model renormalized in the
$\overline{\mathrm{MS}}$ scheme at some high renormalization scale, which is
chosen by the user, are evaluated in perturbation theory as precisely as
possible in two steps. 
First, the initial conditions at the electroweak energy scale are evaluated
from the Fermi constant $G_F$ and the pole masses of the $W$, $Z$, and Higgs
bosons and the bottom and top quarks including the full two-loop threshold
corrections.
Second, the evolution to the high energy scale is performed by numerically
solving the renormalization group evolution equations through three loops.
Pure QCD corrections to the matching and running are included through four
loops. \\
  {\em Solution method:}\\
  numerical integration of analytic expressions \\
  {\em Restrictions:}\\
  {\em Unusual features:}\\
  {\em Additional comments:}\\
  Available for download from URL: \url{http://apik.github.io/mr/}. \\
The \texttt{MathLink} interface is tested to work with
\texttt{Mathematica} 7--9 and, with an additional flag, also with
\texttt{Mathematica} 10 under \texttt{Linux} and with
\texttt{Mathematica} 10 under \texttt{Mac OS X}. \\
  {\em Running time:}\\
  less than 1 second \\

\end{small}


\section{Introduction}
\label{sec:introduction}

The Standard Model (SM) exhibits excellent agreement with an enormous wealth of
experimental data.
It is supposed to describe particle physics from low energies up to
tera-electron-volt scales and beyond.
Moreover, the recent discovery of a Higgs-like particle
\cite{aad:2012tfa,Chatrchyan:2012xdj} makes the SM one of the best candidates
for the theory of the electroweak (EW) and strong interactions.
The absence of clear signals for {\it new physics}, e.g., supersymmetry, at the
LHC leads us to the question:
how far up in the energy scale can the SM be extrapolated?
The answer to this question depends, among numerous factors, on whether the
EW vacuum remains stable at ultimatively high energies or not.
Recent studies of the scalar potential show that, with the Higgs boson mass
being around 125~GeV, the SM can at most become metastable, in the sense that
the lifetime of our EW vacuum exceeds the age of the Universe,
when extrapolated up to the scale of the Planck mass
\cite{Bezrukov:2012sa,Degrassi:2012ry,Buttazzo:2013uya,Bednyakov:2015sca}.

The analysis of the scalar potential at high energies basically requires two
ingredients:
(i) the renormalization group (RG) evolution of the running parameters
and (ii) the initial conditions that relate the latter to the physical
observables.
These initial conditions, which are determined by the so-called threshold
corrections, are usually taken at some lower energy scale, which is typically
of the order of the masses of the weak gauge bosons or the top quark.
For the consistent use of $L$-loop RG evolution, one should take into account
at least $(L-1)$-loop matching.

It is the goal of this paper to introduce the \texttt{C++} program library
\texttt{mr}, which is designed to conveniently perform the above analysis at
the next-to-next-to-leading-order (NNLO) EW level.
This means that we take into account the full two-loop threshold corrections
and the full three-loop RG equations.
In addition, we collect in our program library all the theoretical knowledge
available to date, including the QCD $\beta$ function and mass anomalous
dimensions through four loops.
In some sense, the \texttt{mr} program library extends the ones of
Refs.~\cite{Chetyrkin:2000yt,Schmidt:2012az}, which are restricted to pure
QCD, to the full SM.

This paper is organized as follows.
Section~\ref{sec:setup} contains the relevant definitions and notations
concerning the parametrization of the SM.
Section~\ref{sec:c++-interface} provides a description of the \texttt{mr}
program library. 
Section~\ref{sec:math-interf} explains how to use the \texttt{Mathematica}
interface.
Section~\ref{sec:typical-examples} points the reader towards example programs.
Section~\ref{sec:conclusion} contains our summary.
\ref{sec:a}, \ref{sec:b}, and \ref{sec:methods-rge} list the methods needed for
the on-shell (OS) mass input, running-parameter input, and the RG evolution,
respectively, and\break \ref{sec:example} guides the reader through a typical
example.

\section{The physical setup}
\label{sec:setup}

The SM can be described in terms of the following running parameters defined in
the modified minimal-subtraction ($\overline{\mathrm{MS}}$) renormalization
scheme at the renormalization scale $\mu$:
\begin{equation}
\label{ms_param}
  g_s(\mu), g(\mu), g^\prime(\mu), y_t(\mu), y_b(\mu), \lambda(\mu), v(\mu),
\end{equation}
where $g_s(\mu)$ is the strong gauge coupling, $g(\mu)$ and $g^\prime(\mu)$ are
the EW gauge couplings, $y_t(\mu)$ and $y_b(\mu)$ are the Yukawa couplings
of the top and bottom quarks, $\lambda(\mu)$ is the scalar self coupling, and
the vacuum expectation value $v(\mu)=\sqrt{m_\phi^2(\mu)/(2\lambda(\mu))}$ is
defined through $\lambda(\mu)$ and the Higgs mass parameter $m_\phi(\mu)$,
which both appear in the scalar potential,
\begin{equation}
  \label{eq:potential}
  V(\phi) = -\frac{m_\phi^2}{2}\phi^\dagger\phi + \lambda(\phi^\dagger\phi)^2.  
\end{equation}
Here, the normalization of the quadratic term has been chosen so that
$m_\phi^0=m_H^0$ at tree level.
In our analysis, we retain the Yukawa couplings only for the top and bottom
quarks. 
The contributions due to the Yukawa couplings of the other fermions may be
safely neglected;
those fermions practically only contribute through their interactions with the
gauge bosons.
In addition to the parameters in Eq.~(\ref{ms_param}), one may also consider
the $\overline{\mathrm{MS}}$ masses,
\begin{equation}
\label{ms_masses}
   m_W(\mu), m_Z(\mu), m_H(\mu), m_t(\mu), m_b(\mu).
\end{equation}
It is well known that, to preserve the gauge independence of the
$\overline{\mathrm{MS}}$ masses in Eq.~(\ref{ms_masses}), one has to
systematically include the tadpole contributions (see, e.g., the detailed
discussions in
Refs.~\cite{Fleischer:1980ub,Hempfling:1994ar,Jegerlehner:2003py,%
Kniehl:2004hfa,Jegerlehner:2012kn,Kniehl:2014yia,Kniehl:2015nwa}).

We first introduce the matching relations, which provide the initial conditions
for the RG differential equations.
These relations define the $\overline{\mathrm{MS}}$ parameters at some
low-energy scale $\mu_0$ in terms of appropriate input parameters.
Specifically, our choice of input parameters is
\begin{equation}
  \label{eq:brinput}
\alpha_s(M_Z),G_F,M_W,M_Z,M_H,M_t,M_b,
\end{equation}
where $\alpha_s(\mu)=g_s^2(\mu)/(4\pi)$ is the $\overline{\mathrm{MS}}$
strong-coupling constant, $G_F$ is Fermi's constant, and $M_W$, $M_Z$, $M_H$,
$M_t$, and $M_b$ are the pole masses of the corresponding particles.
At $\mu=M_Z$, $n_f=5$ quark flavors are considered active. 

It is also useful to introduce the $\overline{\mathrm{MS}}$ electromagnetic
gauge coupling via \cite{Kniehl:2015nwa}
\begin{equation}
\label{eq:e}
  \frac{1}{e^2(\mu)} = \frac{1}{g^2(\mu)} + \frac{1}{g^{\prime2}(\mu)},
\end{equation} 
which in turn defines the $\overline{\mathrm{MS}}$ fine-structure constant as
\begin{equation}
\label{eq:a_mu}
  \alpha(\mu) = \frac{e^2(\mu)}{4\pi}.
\end{equation} 

If the input in terms of the OS parameters in Eq.~(\ref{eq:brinput}) is given,
then the $\overline{\mathrm{MS}}$ couplings can be obtained at the 
matching scale $\mu_0$.
The corresponding matching relations are parametrized as
\begin{eqnarray}
  g^2(\mu_0) &=& 2^{5/2} G_F M_W^2 [ 1 + \delta_W(\mu_0) ],
\nonumber
\\
  g^2(\mu_0) + g^{\prime2}(\mu_0) &=& 2^{5/2} G_F M_Z^2 [ 1 + \delta_Z(\mu_0) ],
\nonumber
\\
  \lambda(\mu_0) &=& 2^{-1/2} G_F M_H^2 [ 1 + \delta_H(\mu_0) ],
\nonumber
\\
  y_f(\mu_0) &=& 2^{3/4} G_F^{1/2} M_f [ 1 + \delta_f(\mu_0) ],
\label{eq:tc}
\end{eqnarray}
where $f=t,b$.
In Eq.~(\ref{eq:tc}), $\delta_x(\mu)$ are complicated functions of the OS
parameters in Eq.~(\ref{eq:brinput}) and $\mu_0$, which may be expanded as
perturbation series,
\begin{equation}
  \label{eq:deltax}
  \delta_x(\mu) = \sum_{i,j} \left(\frac{\alpha(\mu)}{4\pi}\right)^i
                   \left(\frac{\alpha_s(\mu)}{4\pi}\right)^j Y^{i,j}_x(\mu).
\end{equation}
The expansion coefficients $Y^{i,j}_x(\mu)$ are generally available for
$i,j=1,2$, which corresponds to two-loop matching.
Beyond that, the pure QCD corrections\footnote{%
In the bosonic cases $x=W,Z,H$, the pure QCD contributions vanish identically,
$Y^{0,j}_x(\mu)=0$.}
are known through four loops and are given by $Y^{0,3}_f(\mu)$ and
$Y^{0,4}_f(\mu)$ \cite{Marquard:2015qpa}.

In addition to Eq.~(\ref{eq:tc}) for the couplings, we also present the
matching relations for the masses, which we write as
\begin{eqnarray}
  m_B^2(\mu_0) &=& M_B^2 [1 + \Delta_B(\mu_0)],
  \qquad B=W,Z,H,
\nonumber\\
  m_f(\mu_0) &=& M_f [1 + \Delta_f(\mu_0)],
  \qquad f=t,b,
  \label{eq:m}
\end{eqnarray}
for the bosons and fermions, respectively.
The functions $\Delta_x(\mu)$ may also be expanded in perturbation series,
\begin{equation}
  \label{eq:mpertfer}
  \Delta_x(\mu) = \sum_{i,j} \left(\frac{\alpha(\mu)}{4\pi}\right)^i
                   \left(\frac{\alpha_s(\mu)}{4\pi}\right)^j X^{i,j}_x(\mu).
\end{equation}
The comments made below Eq.~(\ref{eq:deltax}) also apply to
Eq.~(\ref{eq:mpertfer}).

The pure QCD corrections to $m_t(\mu)$ and $y_t(\mu)$ are evaluated according
to Eq.~(59) in Ref.~\cite{Kniehl:2015nwa} with $n_l=5$ massless quarks and
$n_h=1$ massive quark.
For the pure QCD corrections to $m_b(\mu)$ and $y_b(\mu)$, to take into account
diagrams with top-quark insertions, we treat the four lightest quarks as
massless, but keep both the top and bottom quarks massive through two loops.
The corresponding analytic formula reads \cite{Gray:1990yh,Bekavac:2007tk}:
\begin{eqnarray}
  \label{eq:topInBottomQCD}
                        \delta_{\rm QCD}(\mu) &=&
                        %
                        %
                                                \frac{\alpha_s}{4\pi}\left(-\frac{16}{3} - 4l_{\mu M_b}\right)
                                                %
                                                %
                                                + \left(\frac{\alpha_s}{4\pi}\right)^2\left\{
                                                -\frac{3305}{18}
                                                -\frac{64\zeta_2}{3}
                                                +\frac{8 \zeta_3}{3}
                                                \right. \nonumber\\
                      &&{}-\frac{32 \zeta_2}{3}\ln 2
 + n_h \left(\frac{143}{9}-\frac{32 \zeta_2}{3}\right)
                        + n_l\left(\frac{71}{9}+\frac{16
                        \zeta_2}{3}\right)\nonumber\\ 
                      &&{} + \frac{n_m}{ 9 t^4}\left[72 t^2 + 71 t^4 - 48 t^2
                        H_{0}(t) + 48(1 + t + t^3 + t^4)
                        H_{-1, 0}(t) \right.\nonumber\\
                      &&{}- \left. 48 H_{1,0}(t) - 96 t^4 H_{0, 0}(t) + 
                        48 t (1 + t^2 - t^3) H_{1, 0}(t)\right] \nonumber\\
                      %
                      %
                      &&{}+l_{\mu M_b}\left[-\frac{314}{3} + \frac{52}{9}(n_h+n_l+n_m)\right]\nonumber\\
                      %
                      %
                      &&{}+\left.l_{\mu M_b}^2\left[
                        -14+\frac{4}{3}(n_h+n_l+n_m)
                        \right]
                        \right\},
\end{eqnarray}
where $n_l=4$, $n_h=1$, $n_m=1$ refers to the massive bottom quark, $t=M_b/M_t$,
$l_{\mu M_b}=\ln(\mu^2/M_b^2)$, and
\begin{eqnarray}
  H_{0}(t)   & =& \ln t,\qquad
             H_{-1,0}(t)  = \ln t \ln(1 + t) + \li_2(-t),\nonumber\\
  H_{0,0}(t)  & =& \frac{\ln^2t}{2},\qquad
              H_{1,0}(t)  = -\ln t\ln(1 - t) - \li_2(t)\nonumber\\
\end{eqnarray}
are harmonic polylogarithms as introduced in Ref.~\cite{Remiddi:1999ew}.

Finally, also the relations inverse to Eq.~(\ref{eq:m}), which express the pole
masses in terms of the $\overline{\mathrm{MS}}$ parameters, may be of
interest \cite{Jegerlehner:2001fb,Jegerlehner:2002em,Martin:2014cxa},
\begin{eqnarray}
M_b^2 &=& m_B^2(\mu_0) \left[1 + \overline{\Delta}_B(\mu_0) \right],
\nonumber\\
M_f &=& m_f(\mu_0) \left[1 + \overline{\Delta}_f(\mu_0) \right].
\label{eq:mos}
\end{eqnarray}
The functions $\overline{\Delta}_x(\mu)$ are parametrized similarly to
Eq.~(\ref{eq:mpertfer}), with expansion coefficients
$\overline{X}^{i,j}_x(\mu)$.
They are included in our \texttt{C++} program library as well.

The running of the $\overline{\mathrm{MS}}$ parameters in Eqs.~(\ref{ms_param}) and
(\ref{ms_masses}) is governed by the RG equations,
\begin{eqnarray}
\label{eq:RG1}
  \mu^2\frac{dx}{d\mu^2} &=& \beta_x,\qquad x=g_s,g,g^\prime,y_t,y_b,\lambda,
\\
\label{eq:RG2}
  \mu^2\frac{d\ln x}{d\mu^2} &=& \gamma_x,\qquad x=m_W,m_Z,m_H,m_t,m_b,
\end{eqnarray}
with the respective $\beta$ functions $\beta_x$ and mass anomalous dimensions
$\gamma_x$.
Given the values of the parameters $x(\mu_0)$ at some initial scale $\mu_0$,
Eqs.~(\ref{eq:RG1}) and (\ref{eq:RG2}) allow us to find their values at some
high scale $\mu$.
Since $\beta_x$ and $\gamma_x$ are in turn functions of the $\overline{\mathrm{MS}}$ parameters in
Eq.~(\ref{ms_param}), the RG equations form a system of nonlinear differential
equations to be solved simultaneously.
The functions $\beta_x$ and $\gamma_x$ have been known through four loops in
QCD for a long time
\cite{vanRitbergen:1997va,Chetyrkin:1997dh,Vermaseren:1997fq,Chetyrkin:2004mf,%
Czakon:2004bu}
and have recently been computed through three loops in the full SM
\cite{Mihaila:2012fm,Chetyrkin:2012rz,Mihaila:2012pz,Bednyakov:2012rb,%
Bednyakov:2012en,Chetyrkin:2013wya,Bednyakov:2013eba,Bednyakov:2013cpa}.
In the case of $\beta_\lambda$, even the mixed four-loop correction of
order $\mathcal{O}(g_s^6y_t^4)$ is available
\cite{Martin:2015eia,Chetyrkin:2016ruf}.

\section{\texttt{C++} library \texttt{mr}}
\label{sec:c++-interface}

The \texttt{mr} program library provides the following ingredients:
\begin{itemize}
\item evaluation of the coefficients $Y_x^{i,j}(\mu)$ and $X_x^{i,j}(\mu)$ in
Eqs.~(\ref{eq:deltax}) and (\ref{eq:mpertfer}), respectively, for given input
values of the OS parameters in Eq.~(\ref{eq:brinput}) and, inversely, of the
coefficients $\bar{X}_x^{i,j}(\mu)$ for the given input values of the
$\overline{\mathrm{MS}}$ parameters in Eqs.~(\ref{ms_param}) and
(\ref{ms_masses}),
\item evaluation of the $\overline{\mathrm{MS}}$ couplings and masses according
to Eqs.~(\ref{eq:tc}) and (\ref{eq:m}), respectively, as well as evaluation of
the inverse relations in Eq.~(\ref{eq:mos}),
\item evolution of the $\overline{\mathrm{MS}}$ parameters in the scale $\mu$
using the RG equations in Eqs.~(\ref{eq:RG1}) and (\ref{eq:RG2}).
\end{itemize}

First, in order to use the \texttt{C++} interface, one has to include the
corresponding header file. 
All parts of the program library are placed in the namespace \texttt{mr}:
\begin{lstlisting}
  #include "mr.hpp"
  using namespace mr;
\end{lstlisting}

\subsection{Input parameters}
\label{sec:input}

Before evaluating the coefficients $X_x^{i,j}(\mu)$ and $Y_x^{i,j}(\mu)$, the
input parameter class must be created.
If we wish to use the OS parameters as input, we have to use the class
\texttt{OSinput} with a constructor taking as arguments the five pole masses
$M_b$, $M_W$, $M_Z$, $M_H$, and $M_t$ (in GeV):
\begin{lstlisting}
  //           Mb     MW       MZ       MH      Mt
  OSinput oi(4.4, 80.385, 91.1876, 125.7, 173.2);
\end{lstlisting}
For the user's convenience, we have predefined sets of input parameters from
different editions of the Review of Particle Physics by the Particle Data Group
(PDG) tabulated in enums named \texttt{pdg20xx}, where \texttt{20xx} denotes
the year of the edition.
For example, we may construct an object using the OS input from the 2014 issue
as:
\begin{lstlisting}
  OSinput oi(pdg2014::Mb, pdg2014::MW, pdg2014::MZ, pdg2014::MH, pdg2014::Mt);
\end{lstlisting}
For a detailed description of the \texttt{OSinput} class methods, see
\ref{sec:class-osinput}.

If we are interested in the inverse relations, defining the pole masses in
terms of the $\overline{\mathrm{MS}}$ parameters, we have to use the class
\texttt{MSinput}. 
It constructs an object from the $\overline{\mathrm{MS}}$ masses (in GeV)
(for a detailed description, see \ref{sec:class-MSinput}):
\begin{lstlisting}
  //                          mb      mW       mZ       mH      mt
  mi = MSinput::fromMasses(4.4, 80.385, 91.1876, 125.7, 173.2);
\end{lstlisting}
or from the $\overline{\mathrm{MS}}$ coupling constants:
\begin{lstlisting}
//                           g1     g2     yb     yt     lam    mphi   scale
mi = MSinput::fromCouplings(0.462, 0.648, 0.023, 0.937, 0.126, 131.55, 173.2);
\end{lstlisting}


\subsection{Evaluation of coefficients $X_x^{i,j}(\mu)$ and $Y_x^{i,j}(\mu)$}

Having at hand the \lstinline!OSinput! class instance, we may construct
instances of classes providing the $\overline{\mathrm{MS}}$ masses and
couplings for given input values of the pole masses:
\begin{lstlisting}
bb<OS>    db(oi, oi.MMt());
WW<OS>    dW(oi, oi.MMt());
ZZ<OS>    dZ(oi, oi.MMt());
HH<OS>    dH(oi, oi.MMt());
tt<OS>    dt(oi, oi.MMt());
\end{lstlisting}
Here, the second parameter is the square of the matching scale $\mu_0$;
in the previous example, we have used $\mu_0^2=M_t^2$.
For the $b$, $W$, $Z$, $H$, and $t$ initialization \texttt{bb<OS>}, etc., this
means that we calculate the $\overline{\mathrm{MS}}$ parameters in terms of the
OS ones.

When we wish to calculate the inverse relations in Eq.~(\ref{eq:mos}), we may
specify the input values of the $\overline{\mathrm{MS}}$ masses as:
\begin{lstlisting}
MSinput mi(mb, mW, mZ, mH, mt);
\end{lstlisting}
In contrast to the OS masses, which are denoted by uppercase letters,
$\overline{\mathrm{MS}}$ masses are denoted by lowercase letters.
The input parameters \texttt{mb}, \texttt{mW}, \texttt{mZ}, \texttt{mH}, and
\texttt{mt} implicitly dependent on the scale $\mu$.
When the coefficients $\bar{X}_x^{i,j}(\mu)$ are to be calculated from the
$\overline{\mathrm{MS}}$ masses, the scale $\mu$ at which the latter are
defined has to be specified:
\begin{lstlisting}
bb<MS> xb(mi, mi.mmt());
WW<MS> xW(mi, mi.mmt());
ZZ<MS> xZ(mi, mi.mmt());
HH<MS> xH(mi, mi.mmt());
tt<MS> xt(mi, mi.mmt());
\end{lstlisting}
Note that the scale $\mu$ is common to all the $\overline{\mathrm{MS}}$ masses.
Here, we have chosen $\mu^2=m_t^2(\mu)$.

For each type of particle and set of input parameters, the construction of the
class instances \texttt{db}, \texttt{dW}, \texttt{dZ}, \texttt{dH}, and
\texttt{dt} or \texttt{xb}, \texttt{xW}, \texttt{xZ}, \texttt{xH}, and
\texttt{xt} is the most time-consuming operation.
At this step, the prototypes of all master integrals are evaluated for the
predefined values in the input masses.

After initialization, the expansion coefficients in Eqs.~(\ref{eq:deltax}) and
(\ref{eq:mpertfer}) become available for each particle $x$.
One may refer to the coefficients $X_x^{i,j}(\mu)$ and $Y_x^{i,j}(\mu)$ through
the following methods:
\begin{lstlisting}
long double <particle>::x(size_t a_order, size_t as_order)
long double <particle>::y(size_t a_order, size_t as_order)
\end{lstlisting}
Alternatively the calls \texttt{xij()} or \texttt{yij()} are also available,
where $i$ and $j$ are integers.
For example, the call \texttt{dZ.x(1,1)} is equivalent to \texttt{dZ.x11()},
and the call \texttt{dt.y(0,3)} is equivalent to \texttt{dt.y03()}, and so on.


\subsection{Evaluation of corrections to $\overline{\textrm{MS}}$ coupling
constants and masses}
\label{sec:p2ms}

For the evaluation of the $\overline{\mathrm{MS}}$ parameters from
Eqs.~(\ref{eq:tc}) and (\ref{eq:m}) as implemented in our code, $\alpha_s(\mu)$
and $\alpha(\mu)$ have to be supplied at the chosen scale $\mu$.

Let us first consider $\alpha_s(\mu)$.
We implemented a routine for the evolution of $\alpha_s(\mu)$ to the desired
scale $\mu$, starting from some initial scale $\mu_0$ and the initial value
$\alpha_s(\mu_0)$, which is $\alpha_s(M_Z)$ by default.
Sometimes, however, it is of interest to perform the matching at some different
scale $\mu_0$.
If $\mu_0\geq M_t$, the corresponding decoupling at the top-quark threshold
$M_t$ has to be taken into account. 
This is implemented in our code through order $\mathcal{O}(\alpha_s^3)$
\cite{Chetyrkin:1997sg,Chetyrkin:1997un}.
In the following example, we perform the evolution of $\alpha_s(\mu)$ from the
initial PDG value at $\mu_0=M_Z$ up to the scale $\mu=1000$~GeV crossing the
top-quark threshold:
\begin{lstlisting}
  // Initial scale is oi.MZ() 4-loop running
  AlphaS aS(oi, pdg2014::asMZ, 4);
  // Crossing Mt threshold
  std::cout << aS(1000) << std::endl;
\end{lstlisting}

Let us now turn to $\alpha(\mu)$.
In our approach, it is defined through Eqs.~(\ref{eq:e}) and (\ref{eq:a_mu}).
If we introduce $G_F$ as an additional input parameter, we may express
$\alpha(\mu)$ as \cite{Kniehl:2015nwa}
\begin{eqnarray}
\label{alphamu}
  \alpha(\mu) = \frac{\sqrt{2}G_F M_W^2}{\pi}
    [ 1 + \delta_W(\mu) ] \left[ 1 - \frac{M_W^2}{M_Z^2}\,
 \frac{1 + \delta_W(\mu)}{1 + \delta_Z(\mu)}
     \right].                                                                   
\end{eqnarray}
Since $\delta_Z(\mu)$ and $\delta_W(\mu)$ depend on $\alpha(\mu)$,
Eq.~(\ref{alphamu}) only provides an implicit definition of $\alpha(\mu)$.
The \texttt{mr} program library provides two different classes,
\texttt{AlphaSolve} and \texttt{AlphaGF}, to obtain $\alpha(\mu)$ (see
\ref{sec:class-p2ms}).
Class \texttt{AlphaSolve} uses a numerical solution of Eq.~(\ref{alphamu}),
while class \texttt{AlphaGF} uses a perturbative reexpansion thereof.

Class \texttt{P2MS} allows us to evaluate all the $\overline{\mathrm{MS}}$
couplings and masses at a chosen matching scale $\mu_0$.
To use it, object \texttt{OSinput} has to be constructed first, by one of the
methods described in Section~\ref{sec:input}.
One also has to create the $\alpha_s(\mu)$ object \texttt{AlphaS} as described
above. 

Here is an example of defining two sets of $\overline{\mathrm{MS}}$ parameters,
one with $\mu_0=M_Z$ and the other one with $\mu_0=M_t$, using the same set of
OS input parameters:
\begin{lstlisting}
// Input: Pole masses and Fermi constant in OS scheme
OSinput oi(pdg2014::Mb, pdg2014::MW, pdg2014::MZ, pdg2014::MH, pdg2014::Mt);
// $\overline{\mathrm{MS}}$ QCD coupling for as(Mt) from as(MZ)
AlphaS as(oi);
// Set of all $\overline{\mathrm{MS}}$ parameters at scale Mt
P2MS pMSmt(oi,pdg2014::Gf, as(oi.Mt()), oi.Mt(), order::all);
// Set of all $\overline{\mathrm{MS}}$ parameters at scale MZ
P2MS pMSmZ(oi,pdg2014::Gf, as(oi.MZ()), oi.MZ(), order::all);
\end{lstlisting}
The arguments in the constructor of \texttt{P2MS} include the \texttt{OSinput}
object, the value of $G_F$, the value of $\alpha_s(\mu_0)$, and the scale
$\mu_0$.
The last argument, \texttt{order}, is a mask indicating the orders to be
included in Eqs.~ (\ref{eq:deltax}) and (\ref{eq:mpertfer}) (see classes
\texttt{AlphaSolve} and \texttt{AlphaGF}).

\subsection{RG evolution}

For the evolution of the $\overline{\mathrm{MS}}$ couplings from the initial
scale $\mu_0$ up to the final scale $\mu$, we implemented in our code the full
three-loop results for the SM $\beta$ functions.
The pure QCD corrections to $\alpha_s(\mu)$ at four loops
\cite{vanRitbergen:1997va,Czakon:2004bu,Bednyakov:2015ooa} and to $y_t(\mu)$ and
$y_b(\mu)$ at four \cite{Chetyrkin:1997dh,Vermaseren:1997fq,Chetyrkin:2004mf}
and five \cite{Baikov:2014qja} loops are also included.
Using these $\beta$ functions, we may construct the system of coupled
differential equations, which can be solved numerically.
For its numerical solution, we use the Cash--Karp modification of the
Runge--Kutta method with adaptive grid step as implemented in the
external routine \texttt{OdeInt} \cite{2011AIPC.1389.1586A}.

Instead of $g^\prime(\mu)$, the rescaled constant
\begin{eqnarray}
\label{eq:g1}
   g_1(\mu) = \sqrt{\frac{5}{3}} g^\prime(\mu),
\end{eqnarray}
which corresponds to the normalization familiar from grand unified theories
(GUTs), is sometimes used in the literature.
Following this tradition, we also use $g_1(\mu)$ rather than $g^\prime(\mu)$ in
our code.
For the purpose of the evolution, it is convenient to introduce the following
$\overline{\mathrm{MS}}$ coupling constants related to the
$\overline{\mathrm{MS}}$ parameters in Eq.~(\ref{ms_param}):
\begin{eqnarray}
a_1(\mu)&=&\frac{5}{3}\,\frac{g^{\prime2}(\mu)}{16\pi^2},\qquad
a_2(\mu)=\frac{g^2(\mu)}{16\pi^2},\qquad
a_s(\mu)=\frac{g_s^2(\mu)}{16\pi^2},\nonumber\\ 
a_t(\mu)&=&\frac{y_t^2(\mu)}{16\pi^2},\qquad
a_b(\mu)=\frac{y_b^2(\mu)}{16\pi^2},\qquad
a_\tau(\mu)=\frac{y_\tau^2(\mu)}{16\pi^2},\nonumber\\
a_\lambda(\mu)&=&\frac{\lambda(\mu)}{16\pi^2},
\label{eq:a-couplings}
\end{eqnarray}
where we have chosen the GUT normalization for $a_1(\mu)$ and introduced the
Yukawa coupling $y_\tau(\mu)$ of the $\tau$ lepton.
Then, Eq.~(\ref{eq:RG1}) becomes the system of the seven coupled differential
equations describing the RG evolution of the $\overline{\mathrm{MS}}$ coupling
constants in Eq.~(\ref{eq:a-couplings}), 
\begin{eqnarray}
\partial_t a_1&=&\beta_{a_1}(\mathbf{a}),\qquad
\partial_t a_2=\beta_{a_2}(\mathbf{a}),\qquad
\partial_t a_s=\beta_{a_s}(\mathbf{a}),\nonumber\\
\partial_t a_t&=&\beta_{a_t}(\mathbf{a}),\qquad
\partial_t a_b=\beta_{a_b}(\mathbf{a}),\qquad
\partial_t a_\tau=\beta_{a_\tau}(\mathbf{a}),\nonumber\\
\partial_t a_\lambda&=&\beta_{a_\lambda}(\mathbf{a}),
\label{eq:rge-7eq}
\end{eqnarray}
where $\mathbf{a}$ is a seven-component vector accommodating all the relevant
$\overline{\mathrm{MS}}$ coupling constants in the SM, 
$\mathbf{a}=\{a_1,a_2,a_s,a_t,a_b,a_\tau,a_\lambda\}$,
and $\partial_t$ means differentiation with respect to $t=\ln\mu^2$.
The evolution of $m_\phi(\mu)$ and $v(\mu)$ is controlled by the two
additional differential equations,
\begin{equation}
  \partial_t m_\phi=\beta_{m_\phi}(\mathbf{a}),\qquad
  \partial_t v  =\beta_{v}(\mathbf{a}),
\label{eq:rge-9eq}
\end{equation}
with
  \begin{equation}
    \label{eq:bm-def}
    \beta_{m_\phi}=\frac{dm_\phi}{d\ln\mu^2}=\frac{m_\phi}{2}\gamma_{m_\phi^2},\qquad
    \beta_v=\frac{dv}{d\ln\mu^2}=v\gamma_v,
  \end{equation}
where we have adopted the definitions of $\gamma_{m_\phi^2}$ and $\gamma_v$ from
Refs.~\cite{Bednyakov:2013eba,Bednyakov:2013cpa}, respectively.
The RG evolution of the $\overline{\mathrm{MS}}$ coupling constants according
to Eq.~(\ref{eq:rge-7eq}) is implemented in class \texttt{CouplingsSM} and the
extension by the RG evolution of the two additional $\overline{\mathrm{MS}}$
mass parameters according to Eq.~(\ref{eq:rge-9eq}) in class
\texttt{ParametersSM}.
Both of them are template classes.
Using template parameters as \texttt{<a1,a2,as,at,ab,atau,alam,mphi,vev>}, it
is possible to specify the orders of the $\beta$ functions used for the
solution of the differential equations.

The following example uses three-loop $\beta$ functions for the EW gauge
couplings and parameters, and four-loop $\beta$ functions for the strong and
Yukawa couplings, and the negative value \texttt{-1} at the sixth template
parameter position means that the coupling $y_\tau(\mu)$ is completely
eliminated from the SM:
\begin{lstlisting}
  ParametersSM<3,3,4,4,4,-1,3,3,3> 
              p(a1, a2, as, at, ab, atau, lam, mphi, vev, NG)
\end{lstlisting}
If we are just interested in the solution of the coupled system of differential
equations describing the evolution of the $\overline{\mathrm{MS}}$ coupling
constants in Eq.~(\ref{eq:rge-7eq}), we may use the short form:
\begin{lstlisting}
  CouplingsSM<3,3,4,4,4,-1,3> 
              p(a1, a2, as, at, ab, atau, lam, NG)
\end{lstlisting}
If the object \texttt{P2MS} has already been constructed, as described in
Section~\ref{sec:p2ms} and \ref{sec:class-p2ms}, it is possible to
create a solver using: 
\begin{lstlisting}
  CouplingsSM<3,3,4,4,4,-1,3> p(p2ms, NG)
\end{lstlisting}
Here, \texttt{p2ms} is the previously constructed object of type \texttt{P2MS}.
In all these examples, \texttt{mu0} is the initial scale (in GeV) for the RG
evolution, and \texttt{NG} is the number of fermion generations in the SM, with
default value $\mbox{\tt NG}{}=3$.

The actual RG evolution proceeds when \texttt{operator()} is called for one of
the above-mentioned classes.
Correspondingly, a seven-component vector containing the values of the
$\overline{\mathrm{MS}}$ coupling constants or a nine-component vector
containing also the two $\overline{\mathrm{MS}}$ mass parameters is returned.
If, for example, we are interested in $m_{\phi}(\mu)$ at the scale
$\mu=1000$~GeV, the following sequence of commands has to be entered:
\begin{lstlisting}
  SMCouplings ai = p(1000);
  // using enum couplings for more readable indexing
  std::cout << ai[couplings::mphi] << std::endl;
  // using numerical index, starting from 0
  std::cout << ai[7] << std::endl;
\end{lstlisting}
We refer to \ref{sec:methods-rge} for further details.

The \texttt{mr} program library is not restricted to corrections already
included in it, and it is straightforward to add new higher-order results as
they become available.
One such missing ingredients is the five-loop contribution to the $\beta$
function of the strong coupling.
In the file \texttt{smRGE.cpp} of the source code, this contribution is put to
zero, but may be put to the actual values in the future. 
The four-loop decoupling relations
\cite{Schroder:2005hy,Chetyrkin:2005ia,Kniehl:2006bg}, which will then be
required for consistency, are already implemented.

Recently, the leading EW four-loop contribution to the $\beta$ function of the
strong coupling was analytically calculated in
Refs.~\cite{Bednyakov:2015ooa,Zoller:2015tha}, but slightly different results
were obtained.
In want of a third independent calculation, both results were implemented,
the one of Ref.~\cite{Bednyakov:2015ooa} as default and the one of
Ref.~\cite{Zoller:2015tha} commented out, and may be quickly interchanged if
necessary.

\section{\texttt{Mathematica} interface}
\label{sec:math-interf}

It is possible to compile a \texttt{Mathematica} interface to the \texttt{mr}
program library.
If the \texttt{Mathematica} installation path is unusual, the path to the
development utils \texttt{mprep} and \texttt{mcc} has to be specified at the
configuration step, e.g., as:
\begin{lstlisting}[language=bash]
  ./configure --with-mcc-path=<MATHDIR>/<ARCH>/CompilerAdditions
\end{lstlisting}
Here, \texttt{<MATHDIR>/<ARCH>/CompilerAdditions} is a path to the directory
containing the \texttt{mprep} and \texttt{mcc} utils.
Compilation under Linux works with \texttt{Mathematica} 7--9 and under
\texttt{Mac OS X} with \texttt{Mathematica} 10.
To compile under Linux with \texttt{Mathematica} 10, one has to use the flag
\texttt{--enable-math10}.

When the \texttt{Mathematica} interface is successfully compiled, it is
possible to load it into \texttt{Mathematica}.
If the \texttt{mr} \texttt{MathLink} executable is placed in the same directory
and, for example, we wish to calculate $m_W^2(\mu)/M_W^2$ at scale $\mu=M_t$,
then a typical session is:
\begin{lstlisting}[language=Mathematica]
  Install["mr"];
  (*      Mb   MW   MZ   MH   MT   mu *)
  mmWMMW[4.4, 80, 91, 125, 173, 173]  
\end{lstlisting}
The output, where the user has to supply the values of the
$\overline{\mathrm{MS}}$ coupling constants at scale $\mu$, \texttt{aEW[mu]}
and \texttt{aQCD[mu]}, then reads:
\begin{lstlisting}[language=Mathematica]
  {
    1 + 185.3545315192623037 aEW[173] 
      +  563.188663413438646 aEW[173] aQCD[173]
      - 19207.76644304067414 aEW[173]^2   
  }
\end{lstlisting}
Alternatively, if we are just interested in the set of coefficients, the
command: 
\begin{lstlisting}[language=Mathematica]
  Install["mr"];
  (*  Mb   MW   MZ   MH   MT   mu *)
  Xt[4.4, 80, 91, 125, 173, 173]  
\end{lstlisting}
returns a substitution list for the top-quark coefficients
$\mbox{\tt yT}=Y_t^{i,j}(\mu)$ in Eq.~(\ref{eq:deltax})
and $\mbox{\tt xt}=X_t^{i,j}(\mu)$ in Eq.~(\ref{eq:mpertfer}):
\begin{lstlisting}[language=Mathematica]
  {
    xt[1, 0] -> 108.36879199119430404, 
    yT[1, 0] -> 2.1032409021919717513,
    xt[1, 1] -> -412.09726673991774271, 
    yT[1, 1] -> -79.271276023543769118, 
    xt[2, 0] -> -13725.00152946124629, 
    yT[2, 0] -> 685.749556148969868
  }
\end{lstlisting}
These functions, supplied with the \texttt{Mathematica} expressions for the SM
$\beta$ functions, which are available, for example, as ancillary files to the
arXiv versions of
Refs.~\cite{Bednyakov:2012rb,Bednyakov:2012en,Bednyakov:2013eba},
are sufficient for the analysis.
They are collected in the file \texttt{rge.m} included in the \texttt{mr}
program library.
Complete examples of analyses using the \texttt{Mathematica} interface are
available in the \texttt{Mathematica} notebook distributed with the \texttt{mr}
program library.


\section{Example programs}
\label{sec:typical-examples}

We supply example programs with the code of the \texttt{mr} program library.
For the compilation of the user's program using the \texttt{mr} program
library, one has to type:\\
\lstinline[language=bash]!g++ -o userprog 'pkg-config --cflags --libs mr'  userprog.cpp!\\
A complete example is presented in \ref{sec:example}.

\section{Conclusion}
\label{sec:conclusion}

We presented the \texttt{C++} program library \texttt{mr} with a
\texttt{Mathematica} interface for the evaluation of the
$\overline{\mathrm{MS}}$ couplings of the SM including the full set of two-loop
threshold corrections and their three-loop evolution with high numerical
precision.
The source code including examples is available from the URL\break
\url{http://apik.github.io/mr/}.
A development version of this program library has already been used for the
analysis of the EW vacuum stability in the SM \cite{Bednyakov:2015sca}.

\section{Acknowledgments}
\label{sec:acknowledgments}

We thank A.~V.~Bednyakov for useful discussions and for testing our program
library.
This work was supported in part by the German Federal Ministry for Education
and Research BMBF through Grant No.\ 05H15GUCC1, by the German Research
Foundation DFG through the Collaborative Research Centre No.\ SFB~676
{\it Particles, Strings and the Early Universe: the Structure of Matter and
Space-Time}, by the Heisenberg--Landau Programme, and by the Dynasty
Foundation.

\newpage

\appendix

\section{Methods needed for OS mass input}
\label{sec:a}

\subsection{Class \texttt{OSinput}}
\label{sec:class-osinput}

This class includes the following methods:
\begin{itemize}
\item 
\lstinline!OSinput(long double Mb, MW, MZ, MH, Mt) !\\
as the class constructor,
\item \lstinline!long double MMb(), MMW(), MMZ(), MMH(), MMt() !\\
to get the squared masses $M_b^2$, $M_W^2$, $M_Z^2$, $M_H^2$, and $M_t^2$,
\item \lstinline!long double Mb(), MW(), MZ(), MH(), Mt() !\\
to get the masses $M_b$, $M_W$, $M_Z$, $M_H$, and $M_t$,
\item\lstinline!OSinput setMb(long double), setMW(), setMZ(), setMH(), setMt() !\\
to reset the values of the masses $M_b$, $M_W$, $M_Z$, $M_H$, and $M_t$ in the
OS input already initialized,
\item\lstinline!long double CW(), SW(), CCW(), SSW() !\\
to get the trigonometric functions 
$\cos\theta_w$, $\sin\theta_w$, $\cos^2\theta_w$, and $\sin^2\theta_w$ of the
weak mixing angle $\theta_w$.
\end{itemize}

\subsection{Classes \texttt{bb<OS>}, \texttt{WW<OS>}, \texttt{ZZ<OS>},
\texttt{HH<OS>}, and \texttt{tt<OS>}}
\label{sec:class-xx-OS}

The results for the coefficients $Y_x^{i,j}(\mu)$ and $X_x^{i,j}(\mu)$ in
Eqs.~(\ref{eq:deltax}) and (\ref{eq:mpertfer}), respectively, are organized as
follows:
\begin{equation}
  \label{eq:boson-def}
  X^{i,j}= \mbox{\tt nL}\cdot X_L + \mbox{\tt nH}\cdot X_H 
+ \mbox{\tt boson}\cdot X_B,
\end{equation}
and similarly for $Y_x^{i,j}(\mu)$, where \texttt{nL} and \texttt{nH} are the
numbers of massless and massive fermion generations, respectively, and
\texttt{boson} is a tag for the purely bosonic contributions.
This splitting allows for the extraction of the individual contributions from
the full results.
The default values, corresponding to the case of the SM, read
$\mbox{\tt nL}{}=2$, $\mbox{\tt nH}{}=1$, and $\mbox{\tt boson}{}=1$.
The pure QCD corrections to $m_t(\mu)$ and $y_t(\mu)$ are evaluated using
Eq.~(59) in Ref.~\cite{Kniehl:2015nwa} with $n_l=2\,\mbox{\tt NL}{}+1$ and
$n_h={}\mbox{\tt NH}$ and those to $m_b(\mu)$ and $y_b(\mu)$
using Eq.~(\ref{eq:topInBottomQCD}) with $n_l=2\,\mbox{\tt NL}$ and
$n_h=n_m={}\mbox{\tt NH}$.

The following methods are available:
\begin{itemize}
\item \lstinline!long double x01(size_t nL = 2, size_t nH = 1, size_t boson=1) !\\
  \lstinline!long double x02(size_t nL = 2, size_t nH = 1, size_t boson=1) !\\
  \lstinline!long double x03(size_t nL = 2, size_t nH = 1, size_t boson=1) !\\
  \lstinline!long double x04(size_t nL = 2, size_t nH = 1, size_t boson=1) !\\
for the pure QCD corrections to $m_f(\mu)$ with $f=t,b$ (classes
\texttt{tt<OS>} and \texttt{bb<OS>}),
\item \lstinline!long double y01(size_t nL = 2, size_t nH = 1, size_t boson=1) !\\
  \lstinline!long double y02(size_t nL = 2, size_t nH = 1, size_t boson=1) !\\
  \lstinline!long double y03(size_t nL = 2, size_t nH = 1, size_t boson=1) !\\
  \lstinline!long double y04(size_t nL = 2, size_t nH = 1, size_t boson=1) !\\
for the pure QCD corrections to $y_f(\mu)$ with $f=t,b$ (classes
\texttt{tt<OS>} and \texttt{bb<OS>}),
\item \lstinline!long double x10(size_t nL = 2, size_t nH = 1, size_t boson=1) !\\
  \lstinline!long double x11(size_t nL = 2, size_t nH = 1, size_t boson=1) !\\
  \lstinline!long double x20(size_t nL = 2, size_t nH = 1, size_t boson=1) !\\
for the corrections of orders $\mathcal{O}(\alpha)$,
$\mathcal{O}(\alpha\alpha_s)$, and $\mathcal{O}(\alpha^2)$, respectively, to
$m_B(\mu)$ with $B=W,Z,H$ and $m_f(\mu)$ with $f=t,b$,
\item \lstinline!long double y10(size_t nL = 2, size_t nH = 1, size_t boson=1) !\\
  \lstinline!long double y11(size_t nL = 2, size_t nH = 1, size_t boson=1) !\\
  \lstinline!long double y20(size_t nL = 2, size_t nH = 1, size_t boson=1) !\\
for the corrections of orders $\mathcal{O}(\alpha)$,
$\mathcal{O}(\alpha\alpha_s)$, and $\mathcal{O}(\alpha^2)$, respectively, to
$\delta_x(\mu)$ with $x=W,Z,H,t,b$,
\item \lstinline!long double xgl10(size_t nL = 2, size_t nH = 1, size_t boson=1) !\\
  \lstinline!long double xgl11(size_t nL = 2, size_t nH = 1, size_t boson=1) !\\
  \lstinline!long double xgl20(size_t nL = 2, size_t nH = 1, size_t boson=1) !\\
for the corrections of orders $\mathcal{O}(\alpha)$,
$\mathcal{O}(\alpha\alpha_s)$, and $\mathcal{O}(\alpha^2)$, respectively, in
the gaugeless limit to $m_B(\mu)$ with $B=W,Z,H$ and $m_f(\mu)$ with $f=t,b$,
and
\item \lstinline!long double ygl10(size_t nL = 2, size_t nH = 1, size_t boson=1) !\\
  \lstinline!long double ygl11(size_t nL = 2, size_t nH = 1, size_t boson=1) !\\
  \lstinline!long double ygl20(size_t nL = 2, size_t nH = 1, size_t boson=1) !\\
for the corrections of orders $\mathcal{O}(\alpha)$,
$\mathcal{O}(\alpha\alpha_s)$, and $\mathcal{O}(\alpha^2)$, respectively,
in the gaugeless limit to $\delta_x(\mu)$ with $x=W,Z,H,t,b$.
\end{itemize}

Instead of using \texttt{xij(...)}, one can also use the notation
\texttt{x(i,j,...)}.
For that, the following additional methods are available:
\begin{itemize}
\item \lstinline!long double x(size_t apow, size_t aspow, size_t nL = 2, size_t nH = 1, size_t boson = 1) !
\item \lstinline!long double y(size_t apow, size_t aspow, size_t nL = 2, size_t nH = 1, size_t boson = 1) !
\item \lstinline!long double xgl(size_t apow, size_t aspow, size_t nL = 2, size_t nH = 1, size_t boson = 1) !
\item \lstinline!long double ygl(size_t apow, size_t aspow, size_t nL = 2, size_t nH = 1, size_t boson = 1) !
\end{itemize}

\subsection{Class \texttt{P2MS}}
\label{sec:class-p2ms}

This is the main class to obtain the $\overline{\mathrm{MS}}$ parameters in
terms of the OS input.
Let us first consider $\alpha(\mu)$, which is defined through $G_F$.
Using fixed values of $G_F$ and $\alpha_s(M_Z)$ together with
\lstinline!OSinput!, we may find $\alpha(\mu)$ as the numerical solution of
Eq.~(\ref{alphamu}):
\begin{lstlisting}
    AlphaSolve(const OSinput & in_, double tol_ = 10e-9, 
                const long double & Gf0_ = pdg2014::Gf, 
                const long double & as_ = pdg2014::asMZ, 
                unsigned order_ = 
                order::x01|order::x10|order::x02|
                order::x11|order::x20|order::x03)
\end{lstlisting}
Alternatively, we may solve Eq.~(\ref{alphamu}) perturbatively, which gives
another definition of $\alpha(\mu)$ at fixed order:
\begin{lstlisting}
    AlphaGF(const OSinput & in_, double tol_ = 10e-9, 
            const long double & Gf0_ = pdg2014::Gf, 
            const long double & as_ = pdg2014::asMZ,
            unsigned order_ = 
            order::x01|order::x10|order::x02|
            order::x11|order::x20|order::x03)
\end{lstlisting}
In the above methods, the argument \lstinline!order! is a bit mask with
predefined values to enable or disable the corrections of given orders.
The tolerance parameter \lstinline!tol_! controls the accuracy of the numerical
solution.

Now we turn to the \texttt{P2MS} object, the declaration of which is:
\begin{lstlisting}:
  P2MS<AlphaT>::P2MS(const OSinput & oi_, const long double & Gf_,
                       const long double &  as_,
                       const long double & mu_, 
                       unsigned ord_)
\end{lstlisting}
Here, \texttt{AlphaT} is one of the possible types of solution,
\texttt{AlphaSolve} or \texttt{AlphaGF}, at scale $\mu$.

In the following, we use the following notations for the
$\overline{\mathrm{MS}}$ couplings:
$g_1(\mu)$ as defined in Eq.~(\ref{eq:g1}), $g_2(\mu)\equiv g(\mu)$, and
similarly for the squares of the couplings, $a_1(\mu)$ and $a_2(\mu)$
as defined in Eq.~(\ref{eq:a-couplings}).
The available methods of the \texttt{P2SM} class include the following:
\begin{itemize}
\item \lstinline!long double a1(), a2(), as(), at(), ab(), alam()!\\
to get $a_1(\mu)$, $a_2(\mu)$, $a_s(\mu)$, $a_t(\mu)$, $a_b(\mu)$, and
$a_\lambda(\mu)$,
\item \lstinline!long double g1(), g2(), gs(), yt(), yb(), lam()!\\
to get $g_1(\mu)$, $g(\mu)$, $g_s(\mu)$, $y_t(\mu)$, $y_b(\mu)$, and
$\lambda(\mu)$,
\item \lstinline!long double mphi(), vev()!\\
to get $m_\phi(\mu)$ and $v(\mu)$,
\item \lstinline!MSinput getMSpar()!\\
to construct \texttt{MSinput} at scale $\mu$ as explained
in \ref{sec:class-MSinput},
\item \lstinline!SMCouplings runningCouplings()!\\
to get the vector \texttt{\{g1,g2,gs,yt,yb,ytau,lam,mphi,vev\}}, with the
correction to $y_\tau(\mu)$ being always zero, and
\item \lstinline!SMCouplings ai()!\\
to get the vector \texttt{\{a1,a2,as,at,ab,atau,alam,mphi,vev\}}, with the
correction to $a_\tau(\mu)$ being always zero.
\end{itemize}

\section{Methods needed for $\overline{\mathrm{MS}}$ parameter input}
\label{sec:b}


\subsection{Class \texttt{MSinput}}
\label{sec:class-MSinput}

Here, we explain the methods of the classes needed if the initial input is
given in terms of $\overline{\mathrm{MS}}$ masses or couplings.
To highlight the difference in the construction between masses and couplings,
we do not use constructors, but special functions which return carefully
constructed objects:
\begin{itemize}
\item 
  \begin{lstlisting}
    MSinput fromMasses(long double mb, long double mW, 
                         long double mZ, long double mH, 
                         long double mt)
  \end{lstlisting}
for the construction from the set of $\overline{\mathrm{MS}}$ masses,
\item
  \begin{lstlisting}
    MSinput fromCouplings(long double g1, long double g2,
                             long double yb, long double yt, 
                             long double lam, long double mphi, 
                             long double scale) 
  \end{lstlisting}
for the construction from the set of $\overline{\mathrm{MS}}$ couplings and the
fixing of $\mu_0$.
\end{itemize}  

\subsection{Classes \texttt{bb<MS>}, \texttt{WW<MS>}, \texttt{ZZ<MS>},
\texttt{HH<MS>}, and \texttt{tt<MS>}}

Here, we explain the methods for the calculation of the QCD and EW corrections
defined in Eq.~(\ref{eq:mpertfer}):
\begin{itemize}
\item \lstinline!long double x01(size_t nL = 2, size_t nH = 1, size_t boson=1) !\\
  \lstinline!long double x02(size_t nL = 2, size_t nH = 1, size_t boson=1) !\\
  \lstinline!long double x03(size_t nL = 2, size_t nH = 1, size_t boson=1) !\\
  \lstinline!long double x04(size_t nL = 2, size_t nH = 1, size_t boson=1) !\\
for the pure QCD corrections to $m_f(\mu)$ with $f=t,b$ (classes
\texttt{tt<MS>} and \texttt{bb<MS>}),
\item \lstinline!long double x10(size_t nL = 2, size_t nH = 1, size_t boson=1) !\\
  \lstinline!long double x11(size_t nL = 2, size_t nH = 1, size_t boson=1) !\\
  \lstinline!long double x20(size_t nL = 2, size_t nH = 1, size_t boson=1) !\\
for the corrections of orders $\mathcal{O}(\alpha)$,
$\mathcal{O}(\alpha\alpha_s)$, and $\mathcal{O}(\alpha^2)$, respectively, to
$m_B(\mu)$ with $B=W,Z,H$ and $m_f(\mu)$ with $f=t,b$.
\end{itemize}

\section{Methods needed for RG evolution}
\label{sec:methods-rge}

Each $\beta$ function is expressed in the form
\begin{equation}
  \label{eq:beta-form}
  \beta_i^{(l)}=\sum\limits_{j_1,\ldots,\j_7=0}^{j_1+\cdots+j_7\le
    l+1}a_1^{j_1}\cdots a_7^{j_1}C(j_1,\ldots,j_7),
\end{equation}
where $l$ is a cutoff with respect to the loop order. 
It is possible to fix $l$ for each $\beta$ function separately using template
parameters:
\begin{lstlisting}
    template < int pocoa1, int pocoa2, int pocoas, 
                int pocoat, int pocoab, int pocoatau, 
                int pocolam > 
  \end{lstlisting}
To eliminate a coupling constant from the theory, one may assign a negative
value to the respective parameter.
For example, the pure QCD $\beta$ function at one loop is referred to as:
\begin{lstlisting}
  CouplingsSM<-1,-1,1,-1,-1,-1,-1>
\end{lstlisting}
The constructor of the appropriate object from the $\overline{\mathrm{MS}}$
coupling constants at the initial scale $\mu_0$ for \texttt{NG} fermion
generations reads:
  \begin{lstlisting}
    CouplingsSM(double a1, double a2, double as, 
                  double at, double ab, double atau, 
                  double lam,
                  double mu0_, size_t NG_ = 3)
  \end{lstlisting}
Here, we have used the input values of the $\overline{\mathrm{MS}}$ coupling
constants defined in Eq.~(\ref{eq:a-couplings}). 
The analogous constructor that also includes $m_{\phi}(\mu)$ and $v(\mu)$ reads:
  \begin{lstlisting}
    ParametersSM(double a1, double a2, double as, 
                   double at, double ab, double atau, 
                   double lam, double mphi, double vev, 
                   double mu0_, size_t NG_ = 3)
  \end{lstlisting}
The constructor of the appropriate object from the \texttt{P2MS} object
(see\break \ref{sec:class-p2ms}) reads:
  \begin{lstlisting}
    ParametersSM(const P2MS<T>& pi, size_t NG_ = 3)
  \end{lstlisting}
The operator \texttt{()} is used for the evolution up to the final scale $\mu$:
  \begin{lstlisting}
    SMCouplings operator()(long double mu)
  \end{lstlisting}
This method is implemented for both classes, \texttt{CouplingsSM} and
\texttt{ParametersSM}, and returns the lists
$\{a_1(\mu),a_2(\mu),a_s(\mu),a_t(\mu),a_b(\mu),a_\tau(\mu),a_\lambda(\mu)\}$
and\break
$\{a_1(\mu),a_2(\mu),a_s(\mu),a_t(\mu),a_b(\mu),a_\tau(\mu),a_\lambda(\mu),
m_\phi(\mu),v(\mu)\}$, respectively, at the final scale $\mu$.
The operator:
  \begin{lstlisting}
    SMCouplings AandB(long double mu)
  \end{lstlisting}
returns two lists of length nine with the values of the seven 
$\overline{\mathrm{MS}}$ coupling constants and the two additional
$\overline{\mathrm{MS}}$ mass parameters together with their $\beta$
functions at the final scale $\mu$.
These $\beta$ functions are defined in Eqs.~(\ref{eq:rge-7eq}) and
(\ref{eq:rge-9eq}).

\section{Complete example}
\label{sec:example}

Here, we present an example program, which calculates the
$\overline{\mathrm{MS}}$  coupling constants at the initial scale $\mu_0=M_t$
for given input values of the pole masses and evolves them up to the final
scale $\mu$.
We use three-loop RG evolution for the gauge couplings, the top Yukawa
coupling, and the Higgs self-coupling, and set the tau Yukawa coupling to zero
everywhere.
\begin{lstlisting}
// Example of Pole masses and Gf conversion to
// set of running couplings, running Higgs mass
// term and running vev in MS scheme

#include "mr.hpp"

int main (int argc, char *argv[])
{
  try
    {
      loglevel = logINFO;
      
      // Input: Pole masses and Fermi constant in OS scheme
      OSinput oi(pdg2014::Mb, pdg2014::MW, pdg2014::MZ, pdg2014::MH, pdg2014::Mt);

      // Running QCD coupling for as(Mt) from as(MZ)
      AlphaS as(oi);

      // Set of all running parameters at scale Mt
      P2MS<AlphaSolve> pMSmt(oi,pdg2014::Gf, as(oi.Mt()), oi.Mt(), order::all);
      
      // Initial values for running, input from pole masses
      ParametersSM<3,3,3,3,3,-1,3,3,0> avP2MS(pMSmt);
      
      std::cout << std::setprecision(3);
      
      for (size_t muPow = 3; muPow <= 20; muPow++)
        {

          SMCouplings av = avP2MS(pow(10,2*muPow));
          
          std::cout << " log10(mu) = " << muPow 
                    << " a1   = " << av[couplings::g1]
                    << " a2   = " << av[couplings::g2]
                    << " a3   = " << av[couplings::gs]
                    << " at   = " << av[couplings::yt]
                    << " ab   = " << av[couplings::yb]
                    << " atau = " << av[couplings::ytau]
                    << " alam = " << av[couplings::lam]
                    << " mphi = " << av[couplings::mphi]
                    << " vev  = " << av[couplings::vev] << std::endl;
        }            
    }
  catch (std::exception &p) 
    {
      std::cerr << p.what() << std::endl;
      return 1;
    }
  
  return 0;
}
\end{lstlisting}

\bibliographystyle{elsarticle-num}
\bibliography{mrcpc}
\newpage
\end{document}